\documentclass[aps,prl,reprint, nobibnotes]{revtex4-1}

\usepackage{blindtext}
\usepackage{xcolor}
\usepackage{amsmath}
\usepackage{mathtools}
\usepackage{physics}
\usepackage{soul}
\usepackage[colorlinks=true, linkcolor=blue, citecolor=blue, urlcolor=blue]{hyperref}

\begin{document}

\title{Hydrogen 21cm Constraints on the Photon's Spin Scale}

\author{Aidan Reilly\textsuperscript{1,2}}
\thanks{areilly8@stanford.edu}
\author{Alessandro Russo\textsuperscript{1,2}}
\thanks{arusso00@stanford.edu}
\author{Philip Schuster\textsuperscript{1}}
\thanks{schuster@slac.stanford.edu}
\author{Natalia Toro\textsuperscript{1}}
\thanks{ntoro@slac.stanford.edu}

\affiliation{\textsuperscript{1}SLAC National Accelerator Laboratory, Stanford University, Stanford, CA 94039, USA}
\affiliation{\textsuperscript{2}Department of Physics, Stanford University, Stanford, CA, 94305, USA}

\begin{abstract}
We explore the fundamental but untested possibility that the photon is a continuous spin particle (CSP) with a small but non-zero spin Casimir $\rho$. When $\rho\neq 0$, the familiar polarization modes of the photon transform non-trivially under Lorentz boosts, leading to deviations from familiar QED. Surprisingly, these deviations are strongest at low energy, but smoothly vanish in the $\rho\rightarrow 0$ limit. In this letter, we compute corrections to the hydrogen 21cm transition rate, which is expected to be particularly sensitive given the small hyperfine energy splitting $\omega$. We find deviations from QED $\propto \rho^2 \alpha^2/\omega^2$ at leading order, suggesting experimental constraints $\rho\lesssim 1$ meV. Building on this work, we expect that a range of other atomic, molecular, or condensed matter systems could be used to provide even more stringent tests of $\rho$ in electromagnetic interactions. 
\end{abstract}

\maketitle

\textit{Introduction} - Beyond familiar boost-invariant helicity states, massless particles capable of mediating long-range forces also include a more general class: continuous spin particles (CSP) \cite{Wigner:1939cj}, for which their helicity quantum numbers transform non-trivially under Lorentz boosts. Both massive and massless particles are characterized by a spin Casimir $\rho$, the ``spin scale''. In the massive case $\rho$ has a discrete set of values, whereas in the massless case it is a continuous parameter, hence the name CSP. CSPs possess an infinite tower of helicity states ($h = 0, \pm 1, \pm 2$, ...), which mix under Lorentz transformations in a way controlled by $\rho$ (see e.g. \cite{Schuster:2014hca,Schuster:2023xqa} for a review). Individual helicity modes are only boost invariant when $\rho=0$, as is generally assumed in deriving field theories of interacting scalar, vector, and tensor fields. However, \cite{Schuster:2013pxj,Schuster:2013vpr} gave evidence for three classes of consistent interactions between CSP particles and matter. Each class faithfully reproduces the coupling of matter to massless scalar, vector, or tensor fields in the $\rho\to 0$ limit, and at energies $\gg\rho$. In this limit, all but one primary helicity mode decouples from the matter. Thus, it seems possible that some or all of the forces of the Standard Model are mediated by CSPs with small $\rho$, with a multitude of experimental implications never before considered. This fundamental and exciting possibility is the subject of this letter. 

We focus on the possibility that the familiar photon is a CSP with small spin scale $\rho$. Atomic systems offer a natural setting in which to test this idea, where precision measurements can be used to probe deviations from standard ($\rho=0$) QED predictions. We first explored effects of $\rho\neq 0$ in atomic systems focusing only on electric dipole transitions in \cite{Reilly:2025lnm}, where we derived a general framework for computing CSP photon mediated transitions. Using O(1) constraints on the hydrogen 2s lifetime, \cite{Reilly:2025lnm} found that $\rho \gtrsim 0.1$ eV is excluded. Ultimately, the experimental sensitivity is limited by the fact that CSP effects in a wide range of systems are suppressed by $\rho v/ \omega$, with $v$ the velocity of the matter particle and $\omega$ the CSP energy.

The IR domination of $\rho\neq 0$ effects is what motivates the present work. The most sensitive experimental probes of $\rho$ are those with low energy, and the lowest energy atomic transitions tend to be those that involve only a spin flip. The simplest and smallest of these is the transition between hyperfine split states of the hydrogen ground state, which gives the famous 21cm line \cite{Pritchard:2011xb}. While we did not consider such transitions in \cite{Reilly:2025lnm}, recent theoretical work provides a consistent method to couple a CSP photon to spin-1/2 fermions \cite{newCSPFermion}, enabling this study. We calculate explicit corrections to the 21cm transition probability when $\rho\neq 0$, allowing us to place new, more stringent constraints on the photon's spin scale $\rho \lesssim 1$ meV, and to motivate a range of follow-up studies in atomic and related systems. 

We note that deviations of stellar cooling rates, from the emission of the $h=0,2$ modes of the CSP photon, naively sets a stronger limit of $\rho \lesssim$ O(neV). However, the region near $\rho \sim 1$ meV is close to the regime where these modes are trapped inside stars, while the $h=3$ modes are too weakly coupled to generate significant cooling. Thus, laboratory probes sensitive to the range $\rho \sim 0.1 - 10$ meV are important to consider. 
\\ 

\textit{Theoretical Framework} - We start with a few remarks regarding the observation of the hyperfine transition in atomic systems. Directly observing the hyperfine transition rate in vacuum is impractical, as the expected lifetime is approximately 11 million years. However, stimulated emission can be induced in a cavity, as demonstrated in the hydrogen maser \cite{Ramsey1960, Ramsey1962}.  In this case, the matrix element for the transition remains effectively the same as in vacuum, but is multiplied by a Bose-enhancement factor proportional to the amplitude of the oscillating magnetic field in the cavity, $B_{\rm osc} \propto \sqrt{N_\gamma}$. The cavity is then sensitive to transitions $\ket{e}\ket{N_\gamma} \leftrightarrow \ket{g}\ket{N_\gamma + 1}$, where $\ket{e,g}$ are the excited and ground states of the atom. This is the quantum mechanical mechanism of Rabi oscillations driven by an oscillating magnetic field, $\mathbf{B}_{\rm osc} = B_{\rm osc}\cos(\omega t) \mathbf{e}_{B}$, where $\mathbf{e}_B$ is a directional unit vector. For $\mathbf{e}_B$ aligned the atoms' spin axis, and for $\omega \approx\omega_{eg}$, where $\omega_{\rm eg} = E_e - E_g$ is energy splitting of the atomic states, a transition between the excited and ground state becomes near resonant.  To leading order in the fine-structure constant $\alpha$, the transition rate is controlled by the vacuum matrix element,
 \begin{equation}
    \label{eq: M_eg}
     M_{eg} = \bra{g}H_I(t) \ket{e} = \cos{(\omega t)}\bra{g}H_I'\ket{e},
 \end{equation} 
 where $H_I(t)$ and $H_I'$ are the time dependent and independent  pieces of the interaction Hamiltonian. Under the rotating wave approximation, a matrix element that varies with time in this manner leads to oscillations from an initial excited state to a final ground state with time dependence given by, 
\begin{equation}
\label{eq: rate}
    |P_g(t)|^2 = \frac{\Omega_R^2}{\Omega_D^2 + \Omega_R^2} \sin^2\left(\frac{t}{2} \sqrt{\Omega_D^2 + \Omega_R^2} \right),
\end{equation}
where $\Omega_D = \omega_{\rm eg}-\omega$ is the detuning of the cavity field from the transition frequency, and $\Omega_R = |\bra{g}H_I'\ket{e}|$ is the Rabi frequency. In this experimental setup, CSP corrections to the photon coupling will change the Hamiltonian vacuum transition matrix element, and therefore alter the Rabi frequency relative to the QED expectation. It merits noting that the generation of the background B-field at frequency $\omega \simeq \omega_{eg}$ could lead to $\rho$ dependent deviations at similar order to those of the atomic transition matrix element. However, we expect that such deviations would largely be absorbed into the definition of the field strength when tuning the cavity field prior to running the experiment. While future study of this subtlety is needed, we focus here on the leading order $\rho$ corrections to the vacuum matrix element, and set conservative bounds on this basis.

We use the formalism in \cite{Schuster:2014hca,Schuster:2023xqa,Schuster:2023jgc} and \cite{newCSPFermion} to study CSP interactions with spin-1/2 matter particles. These works employ the quantum worldline formalism, wherein a matter particle's worldline Lagrangian is perturbatively coupled to a background photon field to compute amplitudes (see e.g. \cite{Corradini:2015tik, Strassler:1992zr,Brink:1976sz, Brink:1976uf} for a review, and \cite{Schuster:2023jgc} for its application to CSP QED). As opposed to a quantum field theory, the worldline formalism restricts to an arbitrary but fixed number of input and output matter states, while the massless degrees of freedom are described using quantum fields. In many instances, this formalism renders calculations far simpler than their QFT counterparts. 

The worldline formalism describes space-time spin-1/2 fermions using a Lagrangian with both Grassmann-even and Grassmann-odd degrees of freedom ($z^{\mu}(\tau)$ and $\psi^{\mu}(\tau)$, plus $\psi^5$) that are each functions of a worldline time parameter $\tau$. As described in \cite{Brink:1976sz, Brink:1976uf}, the appropriate free Lagrangian realizes a local worldline supersymmetry that transforms $z^{\mu}$ into $\psi^{\mu}$ and vice versa (a linear transformation for $\psi^5$ is recovered by introducing an auxiliary $z^5$). Upon quantizing the free theory, the wave-function $\ket{\psi}$ for the matter state satisfy Hamiltonian constraints $(p^2-m^2)\ket{\psi}=0$ and $(p_{\mu}\psi^{\mu}+m \psi^5)\ket{\psi}=0$, where $p^{\mu}$ is the momentum conjugate to $z^{\mu}$. Morever, quantization results in the $\psi^{\mu}$ operators satisfying the Dirac algebra $\{\psi^{\mu},\psi^{\nu}\}=2\eta^{\mu\nu}$, permitting the identifications $\psi^{\mu}\to \gamma^5\gamma^{\mu}$ and  $\psi^5 \to \gamma^5$, so that the constraints above become the Klein-Gordon and Dirac equations.  

Using the worldline formalism in familiar QED, the interaction Hamiltonian with a background photon plane wave $A_{\mu}(x)=\epsilon_{\mu}(k)e^{ik\cdot x}+c.c.$ of momentum $k^{\mu}$ takes the form, 
\begin{align}
\label{eq:H_qed}
    H^{QED}_I(t) &= A_h \frac{q}{m} e^{i k \cdot z} \left( p\cdot\epsilon(\mathbf{k})+  (k \cdot \psi)(\epsilon(\mathbf{k}) \cdot \psi) \right),
\end{align}
where we have identified $t=z^0(\tau)$, $q$ is the electric charge of the matter, $m$ is its mass, and the addition by the complex conjugate (to make the Hamiltonian real) is left implicit. We have included a factor $A_h$ to encode the occupation number associated with the background magnetic field. In particular, $A_{h}(\mathbf{k}) = B_{\rm osc}/2\omega$. 

Note that under a gauge transformation $\epsilon_{\mu}\rightarrow \epsilon_{\mu}+\beta k_{\mu}$, the first term transforms by a total $\tau$ derivative (from $p^{\mu}=m\partial_{\tau}z^{\mu}$), while the spin term is invariant because the $\psi^{\mu}$ are anti-commuting. Thus, observables computed with this Hamiltonian are gauge invariant.  When this Hamiltonian is inserted between initial and final spin eigenstates, the $\psi^{\mu}$ fields can be replaced by gamma matrices via the correspondence $\psi^\mu \rightarrow \gamma^5 \gamma^\mu$. For example, for generic spin states $\ket{S_i}$ and $\ket{S_f}$, we have:
\begin{equation}
    \langle S_f | (k \cdot \psi)(\eta \cdot \psi) | S_i\rangle \Rightarrow -k_{\mu} \eta_{\nu} \langle S_f |  \gamma^\mu \gamma^\nu  | S_i \rangle ,
    \label{eq:matrix_el_0}
\end{equation}
where we have anti-commuted the $\gamma^5$ matrices through and used the identity $\gamma^5\gamma^5 = -1$. The position / momentum part of the Hamiltonian just becomes the usual derivative operators on the wave function.

While unfamiliar to most readers, the generalization to CSP interactions with vector correspondence in \cite{Schuster:2023xqa} and \cite{newCSPFermion} is straightforward. For $\rho\neq 0$, the photon is identified with the $h=\pm 1$ modes of a field $\Psi(\eta,x)$ that depends on an auxiliary four-vector $\eta^{\mu}$. Though not precise, the reader can think of the tower of helicity modes of the CSP as encoded in the tower of symmetric tensors that occur in the expansion of $\Psi(\eta,x)=\Psi_0(x)+\Psi_1^{\mu}(x)\eta_{\mu}+...$ in powers of $\eta_{\mu}$. On-shell CSP wave-functions can be decomposed into helicity eigenstates $\Psi_{h,k}(\eta,x)=\Psi_{h,k}(\eta)e^{ik\cdot x}+c.c.$, where $\Psi_{h,k}(\eta)$ is an appropriate eigenfunction, entirely analogous to the polarization tensors $\epsilon_{\mu}(k)$ in the QED case. To first-order in charge $q$, and now with $\rho\neq 0$, the interaction Hamiltonian is, 
\begin{align}
\label{eq:H}
    H_I(t) &= \frac{q}{m} A_h (\mathbf{k}) \int [\bar{d}^4 \eta] \, \Psi_{h,k} (\eta) e^{i k \cdot z} e^{-i \rho \frac{\eta \cdot p}{k \cdot p}} \notag \\
    &\quad \times \sqrt{2}i \left( \frac{k \cdot p}{\rho} +  (k \cdot \psi)(\eta \cdot \psi)\right),
\end{align}
where $[\bar{d}^4\eta]$ is the appropriately regulated $\eta$ space measure as defined in \cite{Schuster:2023xqa}. However, for the purpose of this work, integrations over $\eta^{\mu}$ can be accomplished using the following identity from \cite{Schuster:2023xqa}: 
\begin{equation}
\label{eq: eta identity}
    \int [\bar{d}^4 \eta] \Psi_{h,k}(\eta) \; F(\eta) = \int_{0}^{2\pi}\frac{d\phi}{2\pi}e^{-ih\phi} F(\eta(\phi)),
\end{equation}
where $\eta^{\mu}(\phi) = \epsilon_+^{\mu}(k)e^{i\phi}+\epsilon_-^{\mu}(k)e^{-i\phi}$, and $\epsilon_{\pm}^{\mu}$ are any pair of complex conjugate  transverse polarization vectors. A convenient choice of $\eta$ is to set $\eta^0 =0$. The normalization is fixed by requiring $\eta^{\mu}(\phi)$ to have unit norm. We note that performing a full Legendre transformation on the CSP Lagrangian defined in \cite{Schuster:2023xqa} poses technical challenges, which motivated the path integral methods derived in \cite{Reilly:2025lnm}. However, for the specific case of single photon emission and absorption, we can truncate to leading order in $q$ when computing the interaction Hamiltonian in Eq. \eqref{eq:H} using, 
\begin{equation}  
    p = m\dot{z} + \mathcal{O}(q), \quad \text{and} \quad p_\psi = \dot{\psi} = 0 + \mathcal{O}(q),
\end{equation}  
for the momentum conjugates to $z^{\mu}$ and $\psi^{\mu}$. 
\\

\textit{Results} -
For the 21 cm transition between the hyperfine-split states of hydrogen, the matrix element that enters into Eq. \eqref{eq: M_eg} is, 
\begin{align}
    \label{eq: M_eg hydrogen}
    &M_{eg} = \sum_{h}\frac{ig}{m} \frac{B_{\rm osc}}{2\omega}\int \frac{d\phi_\eta}{2\pi}e^{-ih\phi_\eta} \\
    &\quad \times\bra{\xi_0,S_f} e^{i k \cdot z} e^{-i \rho \frac{\eta \cdot p}{k \cdot p}} \notag \left( \frac{k \cdot p}{\rho} + k_\mu\eta_\nu  \gamma^\mu\gamma^\nu \right)\ket{\xi_0, S_i},
\end{align}
where $\xi_0$ is the ground state wave function of the hydrogen atom, $\ket{S_{i,f}}$ are the initial and final spin states, and we have set  $A_{h}(\mathbf{k}) = B_{\rm osc}/\sqrt{8}\omega$ (note that the $\sqrt{2}$ difference in $A_h(\mathbf{k})$ relative to that in Eq. \eqref{eq:H_qed} comes from normalization differences when defining a B-field in $\eta$ space). 
The two-spin system defined by the electron and proton of the hydrogen atom has a triplet excited state and singlet ground state. In the above, $\ket{S_i}$ can be any of the triplet states, while $\ket{S_f}$ is the singlet. We note that since $\braket{S_f}{S_i}=0$, the first term of Eq. \eqref{eq: M_eg hydrogen} (i.e. the term $\propto k\cdot p/\rho$) does not contribute to the matrix element. Furthermore, the $\gamma$ matrices act only on the spin part of the wave-function, while the position and momentum operators act only on the spatial part of the wavefunction $\xi_0$. We can therefore factorize the relevant terms in the integrand of Eq. \eqref{eq: M_eg hydrogen} as, 
\begin{equation}
    \label{eq: factorized integrand}
    \bra{\xi_0} e^{i k \cdot z} e^{-i \rho \frac{\eta \cdot p}{k \cdot p}}\ket{\xi_0}\times k_\mu\eta_\nu \bra{S_f}\gamma^\mu\gamma^\nu\ket{S_i}.
\end{equation}
It is clear from this form that a non-zero $\rho$ only enters through the spatial wave-function overlap factor in \eqref{eq: factorized integrand}.  Indeed, taking $\rho\to 0$ and the long-wavelength limit, the first factor approaches 1 while the second factor is simply the $\eta$-space formulation (see \cite{Schuster:2023xqa}) of the familiar QED result in the long-wavelength limit. 

Since the 21cm transition is not a relativistic process, for the rest of our treatment, we work at leading order in velocity while still considering all orders in the separate expansion parameter $\rho v / \omega$, which may be large even when $v$ itself is small. In particular, we take $k\cdot z \approx \omega t$ and $k\cdot p \approx \omega m$, which allows us to expand the position space part of the Hamiltonian as,  
\begin{equation}
 H_P = e^{ik\cdot z}e^{-i\rho\frac{\eta\cdot p}{k\cdot p}}  =   e^{i\omega t}e^{i\rho\frac{\vec{p}\cdot \vec{\eta}}{\omega m}} + \mathcal{O}(v),
\end{equation}
recalling that $\eta^0 = 0$. Using this approximation, we can rewrite this contribution to the matrix element as, 
\begin{equation}
    \label{eq: full rho solution}
\langle \xi_0 |H_P | \xi_0 \rangle= e^{i\omega t}e^{-\frac{\abs{\rho}\alpha}{ \omega}}\frac{\abs{\rho}^2\alpha^2 + 3 \abs{\rho} \alpha  \omega + 3 \omega^2}{3 \omega^2},
\end{equation}
where we used ${\vec\eta}^2 = 1$ and worked in momentum space to simplify the calculation. As expected, in the limit that $\rho \to 0$ we recover the simple QED term $e^{i\omega t}$, and in the $\rho\to \infty$ limit the amplitude remains finite. With no dependence on the angular position of $\vec{\eta}$, this $\rho$ dependent correction shows up as an overall constant change to the QED matrix element, but does not affect the allowed helicity of the emitted photon. Only $h=\pm 1$  modes are emitted. We verify this statement in detail by computing the spin dependent piece of Eq. \eqref{eq: M_eg hydrogen} in the Supplementary Material, and show that it matches the standard QED calculation. Putting the above together, and expanding in $\rho \alpha / \omega$ to leading order, the complete matrix element is,  
\begin{align}
\label{eq: M_eg leading order}
    M_{eg} &= \sum_{h=\pm1}\frac{ig}{m}\frac{B_{\rm osc}}{4}\langle \xi_0 |H_P | \xi_0 \rangle  \nonumber \\
    &=  \frac{ig}{m}\frac{B_{\rm osc}}{2}\left(1-\frac{1}{6}\frac{ \abs{\rho}^2\alpha^2}{  \omega^2}\right)\cos{(\omega t)} + \mathcal{O}\left(\frac{ \abs{\rho}^4\alpha^4}{  \omega^4}\right),
\end{align}
where $\tfrac{g}{2m}B_{\rm osc}$ is the QED ($\rho\rightarrow 0$) limit of the Rabi frequency $\Omega_R$. The fact that the lowest order correction to QED shows up at $\mathcal{O}(\rho^2)$ is not surprising, as an odd power of $\rho$ term would show up with an odd number of momentum operators, and thus vanish by parity. 

If one were to use this transition to determine the electron's magnetic moment by defining $H_I = -\vec{\mu}_e\cdot \mathbf{B}_{\rm osc}$, the $\rho$ dependent term in Eq. \eqref{eq: M_eg leading order} would manifest as a correction to the electron's landé-g factor, $g_e$. However, this correction would not be the same for measurements of $\mu_e$ in other systems, and in particular is frequency dependent.
\begin{figure}
    \centering
    \includegraphics[width=1\linewidth]{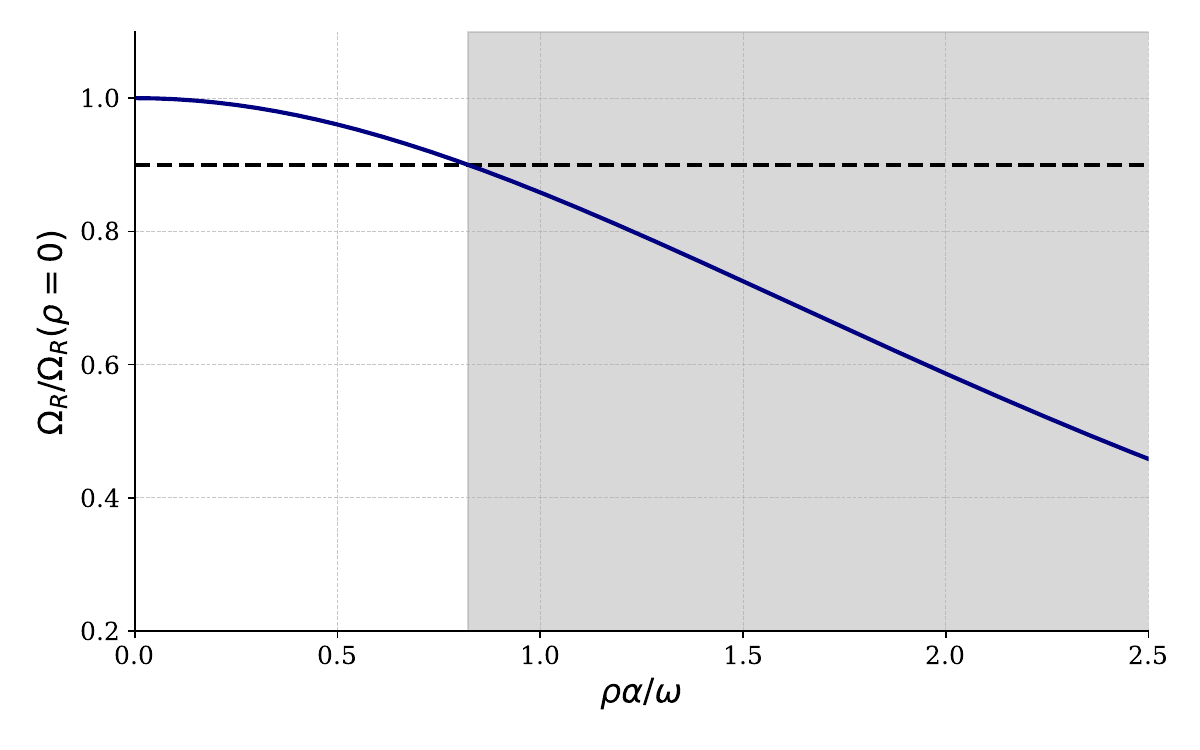}
   \caption{The blue solid line shows the behavior of the Rabi frequency $\Omega_R = |\bra{g}H_I'\ket{e}|$ for $\rho\neq 0$, normalized to the standard case ($\rho = 0$), as a function of $\rho \alpha / \omega$. Based on the data in \cite{Diermaier:2016fsy}, we estimate that $\Omega_R$ can deviate by at most 10\% (indicated by the dashed line) from the standard value. The shaded region represents the excluded parameter space for $\rho \alpha / \omega$.}
    \label{fig:enter-label}
\end{figure}
As the Rabi frequency in Eq. \eqref{eq: rate} is directly proportional to the amplitude $M_{eg}$, any constraint on $\Omega_R$ in a cavity experiment will in turn bound the size of $\rho/\omega$. For example, one can re-purpose the experimental data used to take spectroscopic measurements of the hydrogen hyperfine energy splitting and immediately bound $\rho$. While hydrogen maser experiments offer one method, often additional spectroscopy is carried out via interferometry of the output photons rather than a direct measure of the power and hence transition probability. Therefore, we turn to more modern, ``in-beam" experiments which measure the energy splitting by directly measuring the atomic transition rate as atoms pass through a microwave cavity parallel to a standing magnetic field \cite{Diermaier:2016fsy}. If we take the energy splitting as a known constant, we can re-interpret the data in \cite{Diermaier:2016fsy} to constrain the uncertainty on $\Omega_R$. Although the cavity geometry of \cite{Diermaier:2016fsy} means that the idealized transition probability given by Eq. \eqref{eq: rate} does not directly apply, the $\rho$ dependence will affect the measurement in a similar manner. In this way, we can conservatively say that $\Omega_R$ should not differ from the QED expectations by more than $\sim 10\%$, and hence $\rho\alpha/\omega \lesssim 0.8$. Given the large uncertainties already present, we can safely neglect Zeeman effects in the energy splitting. Using the energy difference $\omega_{\rm hf} = 6.9\times 10^{-6}$ eV \cite{Hellwig1970,Essen1971frequency,Ramsey1990experiments}, we find that
\begin{equation}
    \rho \lesssim 1 \text{ meV}.
\end{equation}
Already, this is two orders of magnitude stronger than the constraint derived in \cite{Reilly:2025lnm}.  
\\

\textit{Discussion And Conclusion} - 
In this letter, we presented a straightforward application of the formalism in \cite{newCSPFermion} to compute the leading order effect that a non-zero spin scale for the photon would have on the dynamics of the hydrogen 21cm transition. We then used this result to derive a constraint on the photon's spin scale $\rho$. In particular, we found that for photons emitted in spin-flip transitions with energy $\omega$, deviations from ordinary QED scale as $\propto \rho^2 \alpha^2 /\omega^2$ in the limit of small $\rho$. This is consistent with the expectation that the novel physics associated with the partner modes of the photon are only important in the IR, at low energy, if $\rho$ is small. At higher energies, the interactions with the partner modes are suppressed -- this is why the otherwise radical possibility that $\rho\neq 0$ for the familiar photon could have gone unnoticed. 

As we noted in the introduction, complementary constraints from stellar cooling are least sensitive in the range $\rho \sim 0.1 - 10$ meV, motivating laboratory tests of QED to decisively probe this range (see \cite{Schuster:2024wjc} for a discussion of CSP thermodynamics). One approach is to improve the precision of the measurements examined in this work, where we currently utilize only an $\mathcal{O}(10\%)$ constraint on the QED prediction. Another approach would be to utilize atomic or molecular transitions with even smaller energy splitting, or to use engineered systems with transitions tuned to be small. An example of such a system might include a one-electron quantum cyclotron, a set-up traditionally used to measure the electron's magnetic moment \cite{Hanneke:2008tm,Fan2023} (note, however, that current experiments operate at cyclotron frequencies higher than that of the 21cm transition). A calculation similar to that performed in this letter could be used to predict the leading order deviations from the QED prediction for the rate of radiation emitted by the electron in this system. Examining changes to the cyclotron frequency, on the other hand, may yield better precision than a rate measurement, but requires solving the full equations of motion, a technically more difficult task. Yet another approach would be to use ``light shining through a wall" style experiments to observe nearby helicity states of the photon \cite{Redondo:2010dp}. Such experiments are limited by $\rho/\omega$ suppression for both the creation and detection of the photon partner modes, but might be able to overcome this limitation with the benefit of extremely low backgrounds and tunable emissions.

In addition to phenomenological studies, many open questions regarding the theory of CSP dynamics remain. Most pressing is that the couplings to matter derived in \cite{Schuster:2023xqa,Schuster:2023jgc} are only strictly valid for on-shell CSP amplitudes. While the current formalism can include off-shell CSPs, it does not provide a unique prediction for their effects. Beyond this limitation, many questions remain about generalizing the Abelian CSP theories explored thus far to non-Abelian theories applicable to the $SU(3)_c\times SU(2)_w$ part of the Standard Model, and also to their confined or higgs phases. Initial progress has been made exploring deviations from General Relativity for the case that gravity has $\rho\neq 0$ \cite{Kundu:2025fsd}, but here too existing calculations are limited to the linear regime, and a full non-linear theory has yet to be developed.

\bibliography{ref.bib}

\section{Supplementary Material} 

\textbf{Spin Calculation and Integral in $\eta$}  

To calculate the spin-flipping part of the Hamiltonian, we first need to define our non-relativistic spin states. In the limit where $v \to 0$ and in the Dirac basis, relativistic spinors take the following form:  

\begin{equation}
    \ket{\uparrow}_{\rm rel} = \begin{pmatrix}
        1\\0\\0\\0
    \end{pmatrix}+\mathcal{O}(v), \quad \ket{\downarrow}_{\rm rel} = \begin{pmatrix}
        0\\1\\0\\0
    \end{pmatrix}+\mathcal{O}(v).
\end{equation}
Using only the upper two indices of these spinors as our non-relativistic bases vectors, we can re-write the spin part of Eq. (\ref{eq: factorized integrand}) as:
\begin{equation}
    k_\mu\eta_\nu \bra{S_f}\gamma^\mu\gamma^\nu\ket{S_i} = i\bra{s_f}\vec{\sigma}\cdot (\vec{k}\times \vec{\eta})\ket{s_i},
\end{equation}
which looks very much like the equivalent QED term, with $\vec{\eta}$ replacing $\vec{A}$. Let us now consider the spin states of interest:
\begin{equation}
\begin{split}
\ket{g} = \ket{F=0, m_F = 0} = \frac{1}{\sqrt{2}}\left(\ket{\uparrow^{\rm e}\downarrow_{\rm p}} - \ket{\downarrow_{\rm e}\uparrow^{\rm p}} \right),\\
\ket{e} = \ket{F=1, m_F = 0} = \frac{1}{\sqrt{2}}\left(\ket{\uparrow^{\rm e}\downarrow_{\rm p}} + \ket{\downarrow_{\rm e}\uparrow^{\rm p}} \right),
\end{split}
\end{equation}
where $F$ is the total spin angular momentum of the atom, $m_F$ is the $\hat{z}$ component of that angular momentum, and the subscripts e and p refer to the spin states of the electron and proton respectively. We have so far only considered coupling of the CSP photon to the electron, an approximation we will continue with as its coupling to the proton will be suppressed by a factor of $m_e/m_p$ relative to the electron, rendering it negligible. We therefore consider only the identity acting on $\ket{\uparrow^{\rm p}},\ket{\downarrow_{\rm p}}$ in the matrix element. This means that terms with opposite proton spins will vanish, and we are left with the computation of 
\begin{equation}
\begin{split}
    \frac{i}{2}\left(\bra{\uparrow^{\rm e}}\vec{\sigma}\cdot (\vec{k}\times \vec{\eta})\ket{\uparrow^{\rm e}} - \bra{\downarrow_{\rm e}}\vec{\sigma}\cdot (\vec{k}\times \vec{\eta})\ket{\downarrow_{\rm e}}\right) = i(\vec{k}\times \vec{\eta})_z .
    \end{split}
\end{equation}
Because $\vec{k}$ and $\vec{\eta}$ are orthogonal, we can define the photon propagation direction to be $\hat{x}$ without loss of generality, so we are left with simply $i\omega \eta_y$. Finally, recalling that the norm of $\vec{\eta}$ is 1, we take our final integration over $\phi_\eta$ according to Eq. \eqref{eq: eta identity}:
\begin{align}
    i\omega\int \frac{d\phi_\eta}{2\pi}e^{-ih\phi_\eta}  \cos{\phi_\eta} = \pm \frac{\omega}{2} \delta_{h\pm1},
\end{align} 
meaning that there only exists support for the helicity one mode. When we sum over photons emitted in the forward and backward direction with helicity $h=\pm1$, we recover a simple factor of $\omega$.

\end{document}